\documentclass{article}

 \usepackage[nonatbib,final]{nips_2017}

\usepackage[utf8]{inputenc} 
\usepackage[T1]{fontenc}    
\usepackage[hidelinks]{hyperref}       
\usepackage{url}            
\usepackage{booktabs}       
\usepackage{amsfonts}       
\usepackage{nicefrac}       
\usepackage{microtype}      
\usepackage{amsmath,graphicx}

\title{NELS - Never-Ending Learner of Sounds}

\author{
   Benjamin Elizalde $^{\ast \dag}$, Rohan Badlani $^{\ast \ddag}$, Ankit Shah $^{\ast \diamond}$, Anurag Kumar$^{\diamond}$,
    Bhiksha Raj$^{ \diamond}$\thanks{All authors contributed equally.}  \\
   $^{\diamond}$LTI, $^{\dag}$ECE, Carnegie Mellon University, Pittsburgh, PA\\
   $^{\ddag}$Department of Computer Science, BITS Pilani, India\\
  \texttt{{bmartin1,aps1,alnu}}@andrew.cmu.edu, \texttt{rohan.badlani@gmail.com} \\
}

\begin{document}

\maketitle
\vspace{-3mm}

\begin{abstract}
Sounds are essential to how humans perceive and interact with the world and are captured in recordings and shared on the Internet on a minute-by-minute basis. These recordings, which are predominantly videos, constitute the largest archive of sounds we know. However, most of these recordings have undescribed content making necessary methods for automatic sound analysis, indexing and retrieval. These methods have to address multiple challenges, such as the relation between sounds and language, numerous and diverse sound classes, and large-scale evaluation. We propose a system that continuously learns from the web relations between sounds and language, improves sound recognition models over time and evaluates its learning competency in the large-scale without references. We introduce the Never-Ending Learner of Sounds (NELS), a project for continuously learning of sounds and their associated knowledge, available on line in \textbf{\href{nels.cs.cmu.edu}{nels.cs.cmu.edu}}. 

\end{abstract}

\vspace{-3mm}

\section{Introduction and Related Work}

\vspace{-2.5mm}

The ability to automatically recognize sounds is essential in a large number of applications such as identifying emergencies in elderly care and patients in hospitals (choking, falling down) \cite{guyot2013two,rocha2018detection}, where monitoring cameras are unwanted due to privacy concerns~\cite{caire2016privacy},  allowing self-driving cars to respond safely to warning sounds and emergency vehicles (ambulance siren)~\cite{DCASE2017challenge}, improving airport and house surveillance, where any number of unusual phenomena have acoustic signatures (alarm, footsteps, glass breaking)~\cite{atrey2006audio}, expanding our interaction with digital assistants through non-verbal communication (clapping, laughing), and analyzing and retrieving video by its content, perhaps the most explored application~\cite{jiang2010columbia, lan2012double,cheng2012sri,schauble2012multimedia,lew2006content}. By the year 2021, a million minutes of videos will be uploaded to the Internet each second; this will constitute 82\% of all consumer traffic \footnote{https://www.cisco.com/c/en/us/solutions/collateral/service-provider/visual-networking-index-vni/complete-white-paper-c11-481360.html}. The ability to recognize sounds in all these recordings is key to organizing, understanding, and exploiting the rapidly growing audio and multimedia data.

In recent years, sound recognition research has focused on curated data and guidelines and although successful and necessary, the literature has under explored the challenges of web audio. Curated audio recordings~\cite{giannoulis2013detection,DCASE2016workshop,DCASE2017challenge,Salamon:UrbanSound:ACMMM:14} have carefully collected and recorded audio as opposed to be mainly recorded in an unstructured manner; have a defined task-oriented set of sound classes as opposed to an unbounded number of sound classes for a wide range of topics; come with a limited set of classes and samples in contrast to an every-day growing number of classes and samples; have rich descriptions of their content in contrast to descriptions that are insufficient, unavailable or wrong. Hence, we should not only test how state of the art sound recognition performs in the web context, but also explore new paradigms to learn from the ever-growing web audio.  

Existing sound recognition systems learn from a finite curated source, so their learning is limited to the scope of the source and the optimization objective and do not improve learning over time. To address these issues, the literature includes never-ending learning architectures that learn many types of knowledge
from years of diverse sources, using previously learned knowledge to improve subsequent learning and with sufficient self-reflection to avoid
learning stagnation, as pointed out by Tom Mitchell~\cite{mitchell2015never}. 

The never-ending paradigm has been employed in ongoing projects, Never-Ending Language Learner~\cite{mitchell2015never} for text and Never-Ending Image Learner~\cite{Chen_2013_ICCV} for images. However, this paradigm is unexplored for sound learning. Examples of tasks related to the paradigm are to learn associations between sounds and language (metadata, ontologies, descriptive terms); continuously grow acoustic vocabularies and improve robustness of sound recognizers; and evaluate the subjectivity of sound recognition in the absence of prior knowledge of the source or generation process.  

We introduce the Never-Ending Learner of Sounds (NELS), a project for large-scale continuous learning of sounds and their associated knowledge by mining the web. Examples of associated knowledge are semantics related to objects, events, actions, places~\cite{lyon2010machine}, cities~\cite{kumar2016audio,elizalde2016city} or qualities~\cite{sager2016audiosentibank,kumar2016discovering}. Our contribution begins with a working framework that serves for audio content indexing and for searching the indexed sounds. Since its inception in 2016, NELS has reported in several research publications discussed in Section~\ref{discussion}, has won the 2017 Gandhian Young Technological Innovation (GYTI) award in India and was a selected abstract in the 2018 Qualcomm Innovation Fellowship.

\vspace{-2.5mm}

\section{NELS Framework}
\label{sec:frmwk}

\vspace{-2.5mm}

In its current form, NELS is a framework that continuously (24/7) crawls audio \& metadata from YouTube videos and creates a content-based index based on a vocabulary of 600 sounds. The sound recognizers were trained from a variety of sources, including web audio itself. NELS also evaluates the quality of the recognition through human feedback. The audio content is indexed combining the crawled metadata, sound recognition predictions and human feedback. As of October 20, 2017, we have crawled over 300 hours of audio corresponding to 4 million video segments of 2.3 seconds. The indexed audio content is available for search and retrieval using our engine in the website.

\begin{figure}[ht]
   \centering
     \includegraphics[width=0.5\textwidth]{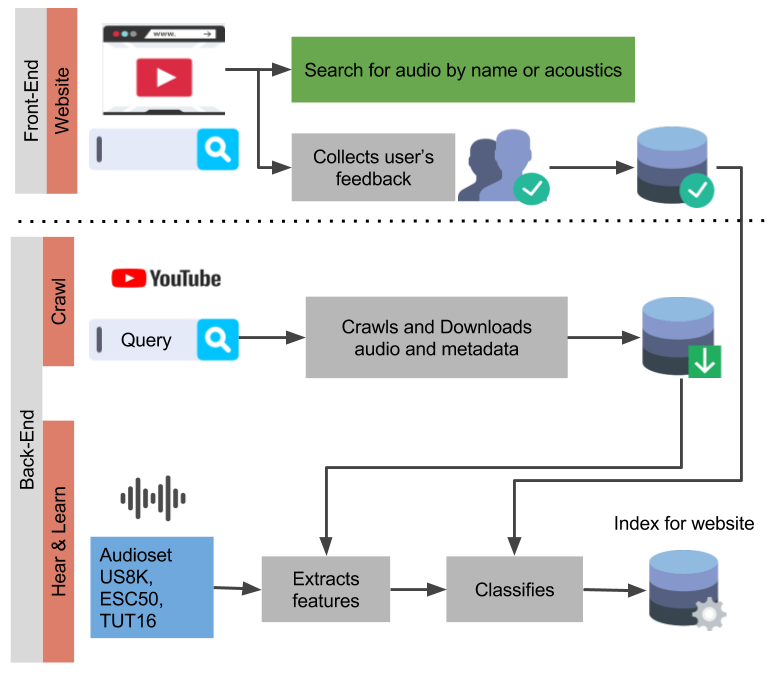}
       \vspace{-2.5mm}
     \caption{The framework serves as a continuously audio content indexer, a sound recognition evaluation and a search engine for the  indexed audio.}
     \vspace{-2.5mm}
     \label{fig:Diagram-Self-learning}
\end{figure}
\vspace{-2.5mm}

\subsection{Crawl}
\label{crawl_fmwk}

\vspace{-2.5mm}
In this module, a search query is used to crawl audio and metadata from YouTube videos. The query corresponds to 605 sound event labels from four different datasets. The video metadata is extracted using the Pafy API\footnote{https://pypi.python.org/pypi/pafy} and corresponds to 12 attributes, such as title, URL and description. The crawled metadata is use to index the audio content.

Although NELS will eventually feed from different sound web archives, we selected YouTube as our first source due to its diversity of sounds and the available metadata associated to them. In contrast to audio-only recordings, collecting audio from videos poses several challenges. YouTube contains massive amount of videos and a proper formulation of the search query is necessary to filter videos with higher chances of containing the desired sound event. Typing a query composed by a noun such as \emph{air conditioner} will not necessarily fetch a video containing such sound event because the associated metadata often corresponds to the visual content; contrary to audio-only archives such as \emph{freesounds.org}. Therefore, we modified the query to be a combination of keywords: ``\textless sound event label\textgreater \hspace{0.1cm}sound", for example,``air conditioner sound". Although the results empirically improved, the sound event was not always found to be occurring and even if it was, sometimes it was present within a short duration, overlapping with other sounds and with low volume. We discarded videos longer than ten minutes and shorter than two seconds because they were either likely to contain unrelated sounds or were too short to be processed. 

\vspace{-2.5mm}
\subsection{Hear \& Learn}
\vspace{-2.5mm}

In this module, we used 605 sound events from four annotated datasets to train classifiers and run them on the crawled YouTube video segments, which are unlabeled. The class predictions are also used to index the audio content.

The framework is being developed so that given a set of guidelines, new datasets could be added seamlessly. NELS should be able to take advantage of existing curated annotations, however dealing with mismatch conditions. The current four datasets are: ESC50, US8k, TUT16 (Task-3) and AudioSet. 
\emph{ESC-50}~\cite{piczak2015environmental} has 50 classes from five broad categories: animals, natural soundscapes and water sounds, human non-speech sounds, interior/domestic sounds and exterior sounds. The dataset consists of 2,000 audio segments with an average duration of 5 seconds each. \emph{The US8K or UrbanSounds8K}~\cite{Salamon:UrbanSound:ACMMM:14} has 10 classes like \emph{gun shot, jackhammer, children playing}. The dataset consists of 8,732 audio segments with an average duration of 3.5 seconds each. \emph{TUT 2016 (Task-3)}~\cite{Mesaros2016_EUSIPCO} has 18 classes like \emph{car passing by, bird singing, door banging} from two major sound contexts namely home context and residential area. The dataset consists of 954 audio segments with an average duration of 5 seconds each. \emph{AudioSet}~\cite{audioset2017} has 527 classes and 2.1 million audio segments with an average duration of 10 seconds each.

The audio from both the datasets and crawled video segments are preprocessed and classified. NELS is meant to be classifier agnostic. We currently follow Convolutional Neural Networks (CNNs) classifier setup described in~\cite{piczak2015environmental}. Recordings are segmented into 2.3 seconds and converted into 16-bit encoding, mono-channel, and 44.1 kHz sampling rate WAV files as in~\cite{piczak2015environmental}. Then, we extracted features comprising of log-scaled mel-spectrograms with 60 mel-bands with a window size of 1024 (23 ms) and hop size is 512. Lastly, the features are used to train multi-class classifiers using CNNs for each of the datasets. 

\vspace{-2.5mm}

\subsection{Website}

\vspace{-2.5mm}

The module is on line in \textbf{\href{nels.cs.cmu.edu}{nels.cs.cmu.edu}} and currently serves for two purposes. First, to evaluate sound recognition using human feedback, which we include as part of the audio indexing. Second, to provide a search engine of the indexed audio content. The goal is to eventually be able to search for audio based on descriptive (subjective) terms, onomatopoeias and acoustic content~\cite{wold1996content}.   

The website provides a search field to capture a term (text-query) for sound searching. The term is mapped to the closest of our sound classes. The mapping uses the tool \textit{word2vec}~\cite{mikolov2013distributed} and a precomputed vocabulary of 400 thousand words called \textit{Glove}~\cite{pennington2014glove}. Word2Vec computes vector representations of the vocabulary words and the text-query. Then, computes cosine similarity between the text-query, the precomputed vocabulary and our list of sounds. Given that our list of sound classes does not necessarily match with all the words of the precomputed vocabulary, we only consider words within a similarity threshold of 0.15, else no results will be retrieved. Additionally, we provide another feature on the website, the user can paste a YouTube video link on a second text field and NELS will yield the dominant sound. 

\begin{figure}[ht]
   \centering
     \includegraphics[width=1\textwidth]{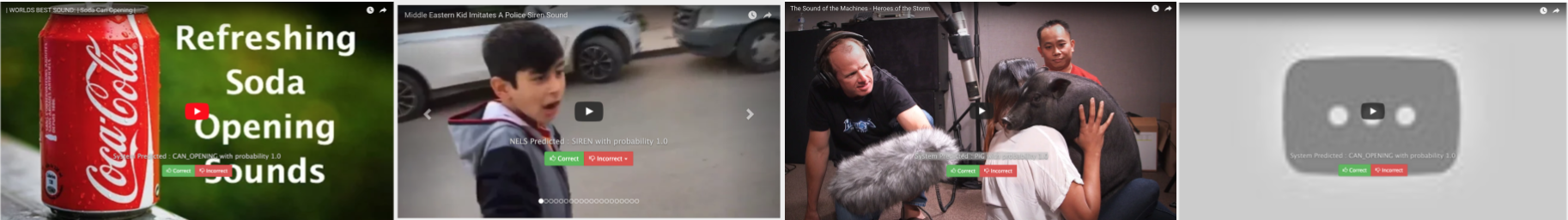}
       \vspace{-2.5mm}
     \caption{Examples of indexed video segments using NELS. First, shows an example where for a \textit{can-opening} sound, the title and images are clearly related. Second, shows an example where a \textit{siren wailing} sound and title are related, but not the visual sound source, which is a child rather than an electronic device. Third, shows an example of \textit{pig-oink} sound, which matches with the visuals, but not with the title and text metadata. Fourth, shows the thumbnail of a video that was indexed, but eventually deleted by the user.}
     \vspace{-2.5mm}
     \label{fig:NELS_examples}
\end{figure}

Each displayed video segment resulting from the text-based search can be evaluated by the user with two button-options, \textit{Correct} or \textit{Incorrect}. That is, whether the human claims that the system's predicted class was present within the segment or not Examples can be seen in Figure~\ref{fig:NELS_examples}.

\vspace{-2.5mm}

\section{Discussion}
\label{discussion}

\vspace{-2.5mm}

In this section we discuss three sound learning challenges where NELS was involved.

\vspace{-2.5mm}

\paragraph{Relation between sounds and language.} Language can describe audio content, be used to search for sounds, and help to define sound vocabularies~\cite{Ellis2018}. However, the relation between sounds and language is still an inchoate topic. To better understand the usage of language for our indexing, we carried two studies. 
  
First, sound recognition results~\cite{giannoulis2013detection,Mesaros2016_EUSIPCO} evidenced how although two sound events: \textit{quiet street} and \textit{busy street} defined audio from streets, the qualifier implied differences in the acoustic content. These kind of nuances can be described with Adjective-Noun Pairs (ANPs) and Verb-Noun Pairs (VNPS)~\cite{sager2016audiosentibank}. We collected one thousand pair-labels derived from different audio ontologies. The audio recordings containing the sound events and their pair-labels were crawled from the collective archive \emph{freesounds.org}. We concluded that despite of the subjectivity of the labels, there is a degree of consistency between sound events and both types of pairs. 

Second, in~\cite{kumar2016discovering} we wanted to identify text phrases which contain a notion of audibility and can be termed as a sound events. We noted that sound-descriptor phrases can often be disambiguated based on whether they can be prefixed by the words ``sound of'' without changing their meaning. Hence, by matching the combination ``sound(s) of <Y>'' where Y is any phrase of up to four words to identify candidate phrases, followed by the application of a rule-based classifier to eliminate noisy candidates, we obtained a list of over 100,000 sound labels. Further, by applying a classifier to features extracted from a dependency path between a manually listed set of acoustic scenes and the discovered sound labels, we were also able to discover ontological relations. For example, forests may be associated with the sounds of \textit{birds singing}, \textit{breaking twigs}, \textit{cooing} and \textit{falling water}.

\vspace{-2.5mm}

\paragraph{Continuous semi-supervised learning of sounds.} NELS should take advantage of existing curated sound datasets and non curated web audio to improve its learning. Previously, semi-supervised self-training approaches have been used to improve sound event recognizers~\cite{han2016semi,shah2016approach}. In~\cite{shah2016approach}, we used an earlier version of NELS. Its classifiers were trained and tested using the US8K dataset consisting of about 8,000 labeled samples for 10 sounds. For re-training, we used 200,000 unlabeled YouTube video segments. Similar to the first paper, but with mismatched data, we achieved about 1.4\% overall precision improvement. Regardless of the improvement, we reached a learning plateau. This could be due to mismatched conditions between training and self-training audio. The initial class bias introduced by the hand-crafted dataset. The use of ambiguous YouTube audio to self-train sound classes. Hence, to learn from the daily growing source of web audio, further exploration is needed.

\vspace{-2.5mm}

\paragraph{Evaluation of the learning quality.} NELS indexes audio content 24/7, but these segments are unlabeled or have weak or wrong labels. Therefore, it is essential to find methods for automatic evaluation of quality in the large-scale. A solution is to include human intervention~\cite{salamon2017scaper}. Hence, our website allows collection of human feedback to asses correctness of sound event indexing. Nevertheless, human feedback may be slow or costly, hence it is important to combine it with other methods that estimate performance in the large-scale. 
  
In~\cite{badlani2018framework}, we used an earlier version of NELS with a recognizer trained on 78 sound events from three different datasets. After, we crawled audio from YouTube videos using the sound event labels from the datasets as the search query. The query was a combination of keywords: ``\textless\emph{sound event label}\textgreater \hspace{0.1cm}sound", for example,``air conditioner sound". Then, we evaluated the highest-40 recognized segments per sound class based on both types of references (ground truth), human feedback and search query. The search query is a summary of the video's metadata describing the whole video, but it was interesting to know to what extent it holds at the video's segment level. Results showed how the performance trends of using both types of references are similar and relatively close with less than an absolute 10\% difference of precision. This trend suggests that the query could be used as a lower-bound of human inspection. In other words, it could serve as a preliminary reference to evaluate sound recognition. Further exploration on this and other associated metadata and multimedia cues could be used as alternative measurements.  

\vspace{-2.5mm}

\bibliographystyle{IEEEbib}
\bibliography{IEEE}

\end{document}